\documentclass[preprint,3p,twocolumn]{elsarticle}

\usepackage{lineno,hyperref}
\modulolinenumbers[5]

\usepackage[utf8]{inputenc}

\usepackage{tabularx}
\usepackage[fleqn]{amsmath}
\usepackage{booktabs}

\usepackage[resetlabels,labeled]{multibib}

\biboptions{numbers,sort&compress}
\newcites{S}{Studies included in the review}

\usepackage{pgfplots}
\pgfplotsset{width=\columnwidth,compat=1.9}

\usepackage{array}
\newcolumntype{L}[1]{>{\raggedright\let\newline\\\arraybackslash\hspace{0pt}}m{#1}}
\newcolumntype{C}[1]{>{\centering\let\newline\\\arraybackslash\hspace{0pt}}m{#1}}
\newcolumntype{R}[1]{>{\raggedleft\let\newline\\\arraybackslash\hspace{0pt}}m{#1}}

\usepackage{xparse}
\let\oriCiteS\citeS

\RenewDocumentCommand{\citeS}{O{} O{} m}{%
  \renewcommand{\citenumfont}[1]{S##1}%
  \oriCiteS[#1][#2]{#3}%
  \renewcommand{\citenumfont}[1]{##1}%
}

\journal{Information and Software Technology}

\begin{document}

\begin{frontmatter}

\title{Learning and Suggesting Source Code Changes from Version History: A Systematic Review}

\author{Leandro Ungari Cayres}
\ead{leandro.ungari@unesp.br}
\author{Bruno Santos de Lima}
\ead{bruno.s.lima@unesp.br}
\author{Rogério Eduardo Garcia}
\ead{rogerio.garcia@unesp.br}
\address{Faculty of Science and Technology, São Paulo State University-UNESP, Presidente Prudente, Brazil}




\begin{abstract}
\textbf{Context:} Software systems are in continuous evolution through source code changes to fixing bugs, adding new functionalities
and improving the internal architecture. All these practices are recorded in the version history, which can be reused as
an advantage in the development process. 
\textbf{Objective:} This paper aims to investigate approaches and techniques related to the learning of source code changes, since the change identification step, learning, and reuse in recommending strategies.
\textbf{Method:} We conducted a systematic review related to primary studies about source code changes. The search approach identified 2410 studies, up to and including 2012, which resulted in a final set of 39 selected papers. We grouped the studies according to each established research question. This review investigates
how source code changes, which were performed in the past of software, can support the improvement of the software
project. 
\textbf{Results:} The majority of approaches and techniques have used repetitiveness behavior of source code changes to identify structural or metrics patterns in software repositories, trough the evaluation of sequences of versions. To extract the structural patterns, the approaches have used programming-by-example techniques to differencing source code changes. 
In quality metrics analysis, the studies have applied mainly complexity and object-oriented metrics.
\textbf{Conclusion:} The main implication of this review is that source code changes as examples, to support the improvement of
coding practice during the development process, in which we presented some relevant strategies to guide each step, since
identifying until the suggesting of source code changes.
\end{abstract}

\begin{keyword}
source code \sep learning code changes \sep history version \sep code metrics \sep code quality
\end{keyword}

\end{frontmatter}



\section{Introduction}

During the software life cycle, several source code chan\-ges are led to fix bugs, make adaptations or even add new functionality, but it can lead to loss of quality and increasing software complexity. 
The software refactoring is a recognized practice for reducing complexity, through small changes, in which source code is restructured without any observable change in external behavior.

However, refactorings are based on simple, widely known and previously cataloged code changes. Refactoring does not take any advantage of the information in the source code history, which does not allow new practices and contributions to be learned. 
On the other hand, there are techniques for learning source code changes based on examples, which allow extracting an edit script of steps to reduce the code change.

In this context, this systematic review of literature aims to identify, evaluate and synthesize quantitative and qualitative studies that investigate or propose approaches and techniques related to the learning of source code changes, since the change identification step, learning, and reuse in recommending strategies.

The article is structured as follows: In Section~\ref{section:background}, we present an overview of source code changes in software development, which details some theoretical bases and previous reviews.
Section~\ref{section:goals} describes the goals of this review and research questions elaborated.
Section~\ref{section:review-method} presents details of the systematic review process, studies selection, data extraction, etc.
Section~\ref{section:results} presents the obtained results and answer the research questions.
Section~\ref{section:discussion} discusses the main benefits and limitations of the extracted evidence.
Section~\ref{section:limitations} presents the limitations of this review.
Section~\ref{section:final-remarks} concludes and provides recommendations for future research on this field.


\section{Background}
\label{section:background}

We first introduce an overview of research studies that are conducted in support of the software development process, mainly related to source code changes, as prediction and fixing bugs, software refactorings, and learning code changes and its recommending.
In second, we summarize the previous systematic reviews related to the main topics of this review.

\subsection{Source code changes in software projects}
Software maintenance is one of the longest stages in the life cycle of a software system.
This phase is composed of activities related to bug fixing, refactoring and adding new functionality, whose changes impact reflects in the internal software quality, resulting in their improvement or degradation~\cite{mens2004survey}.
Any source code change applied may require effort and time to be performed, due to the complexity that the software can acquire, in addition to the propensity to introduce bugs.

The adoption of version control systems has been fundamental in the analysis of software projects since they allow the extraction of source code changes performed in previous versions.
The first approaches have analyzed changes in terms of insertions and deletions of lines of code based on the algorithm proposed by Myers~\cite{myers1986ano}. 
However, these approaches do not adequately fit in the syntax of the programming language.
Chawathe et al.~\cite{chawathe1996change} proposed the use of the abstract syntax tree (AST) in the static analysis of software repositories. This approach describes source code changes in terms of operations on the nodes, according to the hierarchical structure presented.

Overall, the studies focused on corrective and adaptive tasks.
Concerning source code defects, most research studies have focused on the tasks of detection and prediction.
Osman et al.~\cite{osman2014mining} and Hanam et al.~\cite{hanam2016discovering} identified common syntactic patterns of defects and fixes of defects in repositories.
Some research studies have used source code metrics~\cite{nisa2015fault,kaur2016predicting,islam2018characteristics} or code smells~\cite{kaur2016predicting} to construct prediction models.
Liu et al.~\cite{liu2018connecting} mapped each version of the repository based on source code and process metrics to build a metric history for each file, observing behavior regarding defect introduction.
At last, some tools provide automatic support in the prediction of defects~\cite{yuan2013changechecker}.

Towards to source code quality improvements, software refactoring is well known in the software evolution process ~\cite{fowler2018refactoring}.
Fowler~\cite{refactoringCatalog} presented a catalog with the main categories of refactorings, widely known by professionals in the area of Software Engineering.
In the literature, some techniques perform the refactoring process in specified changes types~\cite{dig2006automated,prete2010template,silva2017refdiff,tsantalis2018accurate}.
On the other side, Meng et al.~\cite{meng2011systematic} and Raychev et al.~\cite{raychev2013refactoring} point out these techniques and related development environments support only a limited set of refactorings -- either simple code changes or a sequence of steps for applying more complex refactorings.
To address these shortcomings, Meng et al.~\cite{meng2013lase} and Rolim et al.~\cite{rolim2017learning,rolim2018learning} have presented approaches that use source code changes as examples, to allow the replication of coding practices previously performed in the software repository.

The repetitive and systematic character of source code changes also allows the reuse of the tasks of recommending in software projects.
Nguyen et al.~\cite{nguyen2013study} have pointed out the repetitive tendency of source code changes and defects inside or among projects. 
Source code examples have also been used to suggest/recommend source code changes in programming courses, to support the exercise solution and provide feedback by a tutor~\cite{zhong2018towards}.


\subsection{Summary of previous reviews}

Previous systematic reviews conducted by Breivold et al.~\cite{breivold2010systematic}, Malhotra~\cite{malhotra2015systematic}, Catal and Diri~\cite{catal2009systematic}, Radjenovic et al.~\cite{radjenovic2013software} and, Dallal~\cite{al2015identifying} describe relevant aspects related to source code changes.
These reports investigated some subjects, mainly about software evolution, which includes tasks as identification and prediction of failures (defects) and opportunities to refactoring. We summarize each one of these reviews.

In a corrective view, Breivold et al.~\cite{breivold2010systematic} point out that most papers focus on using a variety of metrics to analyze the evolution of software over time.
Different levels of granularity are used, which result in various perspectives of results.
To support analysis, some resources as comments, structure and quality characteristics of source code, bug tracking, and tools that support data retrieval for evolution analysis are relevant.

Malhotra~\cite{malhotra2015systematic}, Catal and Diri~\cite{catal2009systematic}, Radjenovic et al.~\cite{radjenovic2013software} performed studies related to fault prediction in software projects. 
The main topics highlighted to object-oriented metrics usage, they are widely used and better predictors than complexity and size metrics.  
Additionally, Malhotra~\cite{malhotra2015systematic} points out the most frequent machine learning techniques for software fault prediction were C4.5, Naive Bayes, Multilayer Perceptron, Support Vector Machines, and Random Forest.

At last review, in an adaptive view, Dallal~\cite{al2015identifying} investigates approaches to refactoring activity.
Among the analyzed approaches, the strategies based on quality metrics, precondition and, clustering.
Another relevant point, the studies did not consider the majority part of the refactorings (only 28.8\%) proposed by Fowler~\cite{refactoringCatalog} were not considered in these studies, which limits the process effectiveness.

Overall, none of these reviews investigates approaches to support improvements in the source code, not only software refactoring but also considering source code changes in a general way, presents in software repositories.


\section{Review Objectives}
\label{section:goals}

This systematic review aims to identify the main research studies, techniques, and approaches used in the process of learning source code changes and how to classify such improvements/degradations, based on the resulting impact on quality metrics through the version history.

To identify the points under investigation, we elaborate on the following research questions:
~\\

\textbf{RQ1} -- How are source code changes detected between versions?

\textbf{RQ2} -- How are code changes applied in source code?

\textbf{RQ3} -- How to identify common patterns between source code changes?

\textbf{RQ4} -- How can the impact of source code change be evaluated between improvement or degradation?

\textbf{RQ5} -- How are suggestions/recommending of source code changes are provided to the user?
~\\

Overall, it is intended to investigate all the processes related to modifications in the source code of software projects.
In our knowledge, no previous systematic reviews were published involving all these concepts and new approaches in a unified study.

\section{Review Method}
\label{section:review-method}

\subsection{Protocol development}
We follow the guidelines proposed by Kelle et al.~\cite{keele2007guidelines} and Nakagawa~\cite{nakagawa2017revisao} in conducting of this review, which both points out a similar sequence of activities as follow: the development of protocol, definition of inclusion and exclusion criteria, search of relevant studies, study quality assessment, data extraction, and synthesis.

\begin{figure}
\includegraphics[width=\columnwidth]{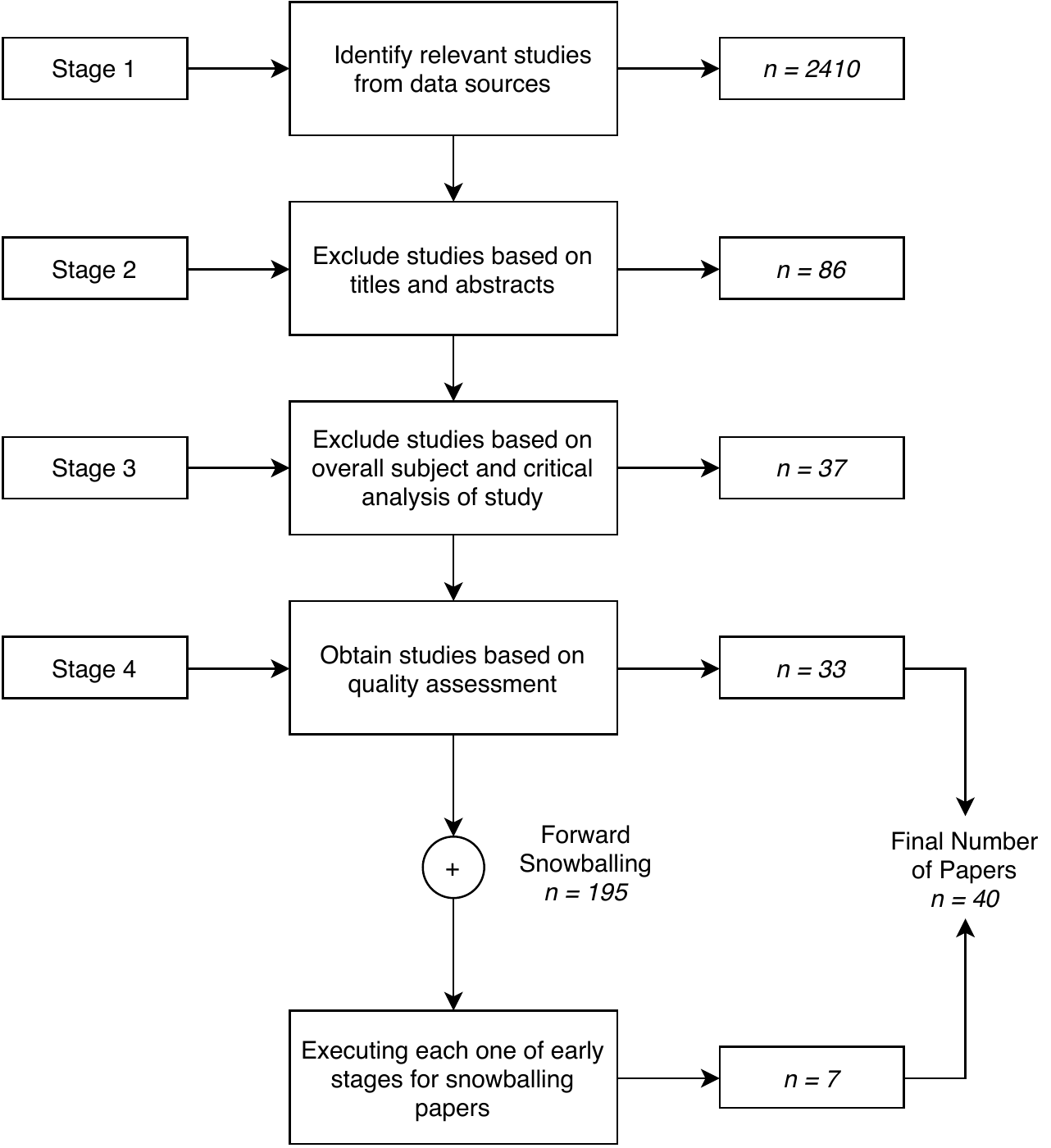}
\caption{Process of studies selection.}
\label{figure:selection-process}
\end{figure}

The search strategy embraced some electronic databases and search engines. The list is presented in Table~\ref{table:data-source}. 
From these data sources, we identified and filtered a set of papers through a selection process by stages, which is presented in Figure~\ref{figure:selection-process}.

\begin{table*}
\caption{Data sources of systematic review}
\label{table:data-source}
\begin{tabularx}{\textwidth}{l|l|l}
\toprule
\textbf{Data source} & \textbf{Online address} & \textbf{Category}\\
\midrule
\emph{IEEE Explore} & \url{https://ieeexplore.ieee.org/Xplore/home.jsp} & Electronic Database\\
\hline
\emph{ACM Digital Library} & \url{https://dl.acm.org/} & Electronic Database and Search Engine\\
\hline
\emph{Engineering Village} & \url{http://www.engineeringvillage.com} & Search Engine\\
\hline
\emph{Science Direct} & \url{https://www.sciencedirect.com/} & Electronic Database\\
\bottomrule
\end{tabularx}
\end{table*}

\begin{figure}
\begin{align}
&~code~transformation\\
&~code~change\\
&~code~edit\\
&~version\\
&~improvement\\
&~degradation\\
&~learning\\
&~code~history\\
&~code~pattern
\end{align}
\caption{List of keywords.}
\label{figure:keywords}
\end{figure}

In Stage 1, we defined a list of keywords (Figure~\ref{figure:keywords}), to identify relevant articles to the systematic review.
These relevant terms were derived and organized using boolean operators (AND and OR) to compose the following search string applied in data sources (Figure~\ref{figure:search-string}):

\begin{figure}
\begin{align*}
&~(code~OR~source~code~OR~code~pattern)~AND~\\
&~(edit~OR~change~OR~transformation)~AND~\\
&~((learning~OR~(history~OR~version)))
\end{align*}
\caption{Search string created in systematic review.}
\label{figure:search-string}
\end{figure}

A result set of 3046 entries was returned, which was composed of 2410 non-duplicated citations.

In Stage 2, these citations were entered in a developed tool to support the process of the systematic review, which lists all studies from BibTeX entry files.
After that, each study entry was analyzed upon the title and abstract of the paper, to identify whether the study is related to the topics of review, based on acceptance and exclusion criteria.
A study was included whether it is compatible at least with acceptance criteria, on the other hand, it was discarded whether it was related to one or more exclusion criteria, which are presented in~\ref{appendix:criteria}.
This systematic review included qualitative and quantitative research studies, only written in English and published since 2012.
This stage has ended up with a remaining total of 86 papers.

In the next stage, these studies were analyzed again upon acceptance and exclusion criteria, but considering a full-text analysis, not only the title and abstract as conducted in the previous step.
As a result of this process, the total number was reduced to 37 papers.

At the last stage, the remaining studies passed through an assessment quality through a form, to verify aspects related to rigor, credibility, and relevance, and at last, to assign a score. 
In Section~\ref{section:quality-assessment} presents the quality assessment process more detailed.
From this evaluation, only in studies whose score was above a minimum threshold, the data extraction process was applied to summarize the main topics of each study and produce results presented in this review.
The final number of studies was reduced to 33.

Additionally, we applied a forward snowballing search to obtain the most recent studies, which were not selected previously.
Beginning from 195 papers, they also passed through the same sequence of stages, to filter the correspondence of subjects and quality of each one, to be added to the final set of review studies (\emph{n = 33 + 7 = 40}).

A pair of researchers performed each stage of the process (1 -- 4) and forward-snowballing, in which an experimenter evaluated the studies and a reviewer to validate previous results.

\subsection{Study quality assessment}
\label{section:quality-assessment}

We performed a quality assessment upon 45 studies, in which 37 and 8 studies were from the review and snowballing process respectively.
Each study was evaluated based on 11 criteria form presented in Table~\ref{table:quality-criteria} (details explained in~\ref{appendix:quality-criteria}).

\begin{table}
\caption{Quality criteria}
\label{table:quality-criteria}
\begin{tabular}{L{.9\columnwidth}}
\toprule
1. Is the article based on scientific research?\\
2. Are the goals clearly defined?\\
3. The study context was clearly presented?\\
4. Is the approach/strategy developed clearly described?\\
5. Did the researchers analyze the advantages/disadvantages/limitations of the approach/strategy?\\
6. How much relevant are the research questions?\\
7. Is data study used in real applications or just experimental context?\\
8. Were the results obtained and analyzed clearly described?\\
9. Was the influence of the researchers evaluated on the results?\\
10. Is there any discussion of the results?\\
11. Were relevant contributions founded in the study?\\
\bottomrule
\end{tabular}
\end{table}

We evaluated the studies on all quality criteria, to assign a score based on the conformity in a range of 0 -- 1 (0:~\emph{Non-acceptable}; 0.5:~\emph{Weakly-acceptable} and 1:~\emph{Strongly-acceptable}).
The final score of the study was obtained by the sum of individual scores.
Only studies, whose score is greater or equal than 6, were accepted to the data extract process (the maximum score possible is 11).

\begin{figure}
\begin{tikzpicture}
\begin{axis}[
    xlabel={Number of study},
    ylabel={Score},
    xmin=0, xmax=45,
    ymin=0, ymax=12,
    xtick={0,9,18,27,36,45},
    ytick={0,2,4,6,8,10,12},
    legend pos=south west,
    ymajorgrids=true,
    grid style=dashed,
]
 
\addplot[
    color=blue,
]
coordinates {
(1,6.5)(2,6.5)(3,8)(4,8)(5,3.5)(6,3.5)(7,9)(8,3.5)(9,8)(10,11)
(11,9)(12,7.5)(13,7)(14,8)(15,9.5)(16,8.5)(17,8)(18,8.5)(19,10)(20,7.5)
(21,10)(22,8)(23,9.5)(24,8)(25,10)(26,8.5)(27,7.5)(28,10)(29,8.5)(30,3)
(31,10)(32,9.5)(33,9)(34,8.5)(35,10.5)(36,8)(37,7.5)(38,10)(39,9.5)(40,10.5)
(41,9)(42,9)(43,10.5)(44,8)(45,6)
};
\addlegendentry{Researcher}
\addplot[
    color=red
]
coordinates {
    (1,7)(2,7.5)(3,9)(4,8)(5,3.5)(6,3.5)(7,10)(8,4.5)(9,7.5)(10,11)
    (11,10.5)(12,9)(13,9.5)(14,8.5)(15,7)(16,9)(17,10)(18,8.5)(19,11)(20,7)
    (21,11)(22,7)(23,9.5)(24,6.5)(25,10)(26,10)(27,7.5)(28,11)(29,9)(30,3)
    (31,11)(32,10.5)(33,9)(34,11)(35,11)(36,11)(37,6)(38,11)(39,10)(40,11)
    (41,7.5)(42,7)(43,11)(44,6.5)(45,5)
};
\addlegendentry{Reviewer}
\end{axis}
\end{tikzpicture}
\caption{Scores of quality assessment process.}
\label{figure:plot}
\end{figure}
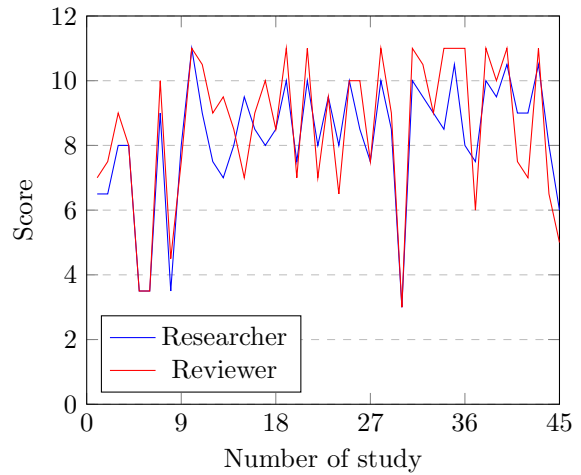

Figure~\ref{figure:plot} presents the assigned scores of each one of the papers by researcher and reviewer. 
The scores refer to all 45 evaluated studies.

The set of criteria was formulated to evaluate the study quality upon the reporting of context, objectives, applied approach or technique, results, and contributions.

We first defined criteria 1--3 to evaluate the study in an overview, to identify the purpose and context. Second, criteria 4--5 are concerned about the approach, evaluation or technique performed in the study and its respective effects. Criteria 6–7 measure the quality of the study evaluation process, and, at last,  8–9 criteria intended to describe and summarize results.

After the evaluation process, we selected all papers which achieved the minimum required score to compose this review. Table~\ref{table:mapping} presents the mapping between the numbering of each paper in the score plot (Figure~\ref{figure:plot}) and their respective referencing in this review. 

The papers which did not achieve the requirements received the label ``Non-accepted''.

\begin{table}
\caption{Mapping between numbering and referencing of papers.}
\label{table:mapping}
\begin{tabular}{L{.27\columnwidth}|L{.27\columnwidth}|L{.27\columnwidth}}
\toprule
1~~\citeS{Skitsu2013detecting} & 2~~\citeS{Syuan2013changechecker} & 3~~\citeS{Sdotzler2016move}\\
4~~\citeS{Slin2016empirical} & 5~{\small Non-accepted} & 6~{\small Non-accepted}\\
7~~\citeS{Snguyen2016using} & 8~{\small Non-accepted} & 9~~\citeS{Skaur2016predicting}\\
10~\citeS{Sfrick2018generating} & 11~\citeS{Snguyen2013study} & 12~\citeS{Smolderez2017mining}\\
13~\citeS{Ssantos2017recommending} & 14~\citeS{Sstevens2017extracting} & 15~\citeS{Smartinez2013automatically}\\
16~\citeS{Smeng2013lase} & 17~\citeS{Ssilva2017refdiff} & 18~\citeS{Sosman2014mining}\\
19~\citeS{Srolim2017learning} & 20~\citeS{Snisa2015fault} & 21~\citeS{Sliu2018connecting}\\
22~\citeS{Sjacobellis2014cookbook} & 23~\citeS{Sfalleri2014fine} & 24~\citeS{Shigo2014mpanalyzer}\\
25~\citeS{Snegara2014mining} & 26~\citeS{Sdotzler2017more} & 27~\citeS{Ssharma2015post}\\
28~\citeS{Snguyen2016api} & 29~\citeS{Smolnar2017discovering} & 30~{\small Non-accepted}\\
31~\citeS{Shead2017writing} & 32~\citeS{Stsantalis2018accurate} & 33~\citeS{Sraychev2013refactoring}\\
34~\citeS{Skessentini2017context} & 35~\citeS{Shanam2016discovering} & 36~\citeS{Smaruyama2016slicing}\\
37~\citeS{Sdagit2013identifying} & 38~\citeS{Sde2018imprecisions} & 39~\citeS{Sliu2018closer}\\
40~\citeS{Szhong2018towards} & 41~\citeS{Shigo2015toward} & 42~\citeS{Ssong2018systematic}\\
43~\citeS{Sislam2018characteristics} & 44~\citeS{Sphothilimthana2017high} & 45~{\small Non-accepted}\\
\bottomrule
\end{tabular}
\end{table}

\subsection{Data extraction}

In each of 43 studies, we applied a form (see~\ref{appendix:form-data-extract}) to extract the relevant topics of each study to compose the result of this review. The main focus of this step is to summarize the selected studies, to answer the questions under investigation and propose new contributions.

We first elaborated a list of attributes to be extracted from remaining studies, this selection has considered previously known studies and relevant systematic reviews (Dyb{\aa} and Dings{\o}yr~\cite{dybaa2008empirical}, Marçal et al.~\cite{marccal2016techniques}).
After that, some adjusting was applied based on the features of context and subject of selected studies, to improve the quality of extraction.

\section{Results}
\label{section:results}

In this section, we describe the obtained results from selected primary studies (PS) in the review (see~\ref{appendix:included-studies}). 
Firstly, we detail some aspects of selected studies, in which we present an overview of the distribution over the years of papers (in Section~\ref{section:publication-year}) and its sources (in Section~\ref{section:publication-source}).
After that, in Sections~\ref{section:rq1},~\ref{section:rq2},~\ref{section:rq3},~\ref{section:rq4} e~\ref{section:rq5}; we answer each one of the research questions.

\subsection{Publication year}
\label{section:publication-year}

\begin{figure}
\begin{tikzpicture}
\begin{axis}[
    xlabel={Year of Publication},
    ylabel={Number of Papers},
    xmin=2011, xmax=2020,
    ymin=0, ymax=12,
    xtick={2012,2014,2016,2018},
    ytick={0,2,4,6,8,10,12},
    ymajorgrids=true,
]
 
\addplot[
    mark=*, mark options={fill=white}
]
coordinates {
(2012,0)(2013,7)(2014,5)(2015,3)(2016,7)(2017,10)(2018,7)(2019,0)
};
 
\end{axis}
\end{tikzpicture}
\caption{Distribution of papers by year.}
\label{figure:plot-year}
\end{figure}
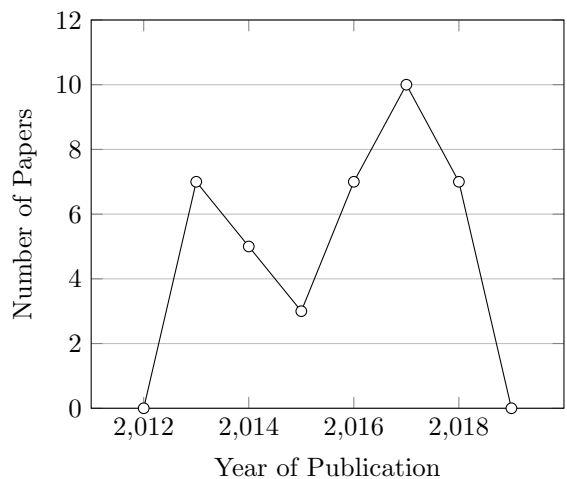

Figure~\ref{figure:plot-year} presents the distributions of papers from the year 2012 to 2019. For the year 2019, the review process has started in January and none study has been published yet.
The most expressive year was 2017 with a total of 10 papers, which supports the last 3 years' growth tendency (accumulative percent of 61\% of studies).

\subsection{Publication source}
\label{section:publication-source}

The number of selected studies per publication source is presented in Table~\ref{table:publication-sources}.
Only sources having two or more papers are highlighted (57\%), others 17 sources have only one paper (43\%).

The majority of the main sources are conferences related to ACM and IEEE, whose research subjects are maintenance, evolution, mining repositories and software engineering. 

\begin{table*}[ht]
\caption{Most important publication sources}
\label{table:publication-sources}
\begin{tabularx}{\textwidth}{p{.8cm} p{9cm} C{1.7cm} C{2.1cm} C{3cm}}
\toprule
\textbf{Rank} & \textbf{Source} & \textbf{Number} & \textbf{Proportion (\%)} & \textbf{Cumulative proportion (\%)}\\
\midrule
1 & IEEE International Conference on Software Maintenance and Evolution (ICSME) & 5 & 13 & 13
\\
2 & IEEE/ACM International Conference on Software Engineering (ICSE) & 5 & 13 & 26
\\
3 & IEEE/ACM International Conference on Automated Software Engineering (ASE) & 4 & 10 & 36
\\
4 & International Conference on Mining Software Repositories (MSR) & 3 & 8 & 44
\\ 
5 & IEEE International Conference on Software Analysis, Evolution and Reengineering (SANER) & 3 & 8 & 52
\\
6 & Intl. Conference on Advances in Computing, Communications and Informatics (ICACCI) & 2 & 5 & 57
\\\bottomrule
\end{tabularx}
\end{table*}

\subsection{Identifying source code changes between versions (RQ1)}
\label{section:rq1}

Source code changes are identified for different purposes, as detecting, predicting and fixing bugs~\citeS{Syuan2013changechecker, Snisa2015fault,Shanam2016discovering,Sislam2018characteristics}, analysis of code quality~\citeS{Skaur2016predicting,Sliu2018connecting,Ssharma2015post}, applying improvements and refactorings~\citeS{Ssilva2017refdiff,Stsantalis2018accurate,Sraychev2013refactoring}, suggesting code changes~\citeS{Ssantos2017recommending,Sjacobellis2014cookbook,Skessentini2017context}, or until fixing a programming lesson~\citeS{Shead2017writing,Sphothilimthana2017high}, and others. 

During analysis, a source code change is extracted from a pair of consecutive versions. According to Falleri et al.~\citeS{Sfalleri2014fine}, a code change was represented in terms of insertions/deletions of lines, based on the algorithm proposed by Myers~\cite{myers1986ano}.
However, this approach does not adequately fit in the source code, which promotes some imprecision. 
Since the study of Fluri et al.~\cite{fluri2007change}, the techniques are using an AST differencing approach, which defines source code changes based on operations over AST nodes between source code entities before and after modifications.

The AST differencing approach is composed of two phases:  mapping, and edit-script generating.
All identified PSs use the optimized algorithm of Chawathe et al.~\cite{chawathe1996change} in the generating task.
So, the PSs~\citeS{Sfalleri2014fine,Sdotzler2016move,Sfrick2018generating} focus on the improvement of the mapping phase, fixing the imprecision of linking between before/after modification entities, to produce edit scripts shorter and more understandable.
In this context, Guillermo et al.~\citeS{Sde2018imprecisions} investigated the state-of-art differencing approach GumTree~\citeS{Sfalleri2014fine} performance under a set of repositories and identified several imprecision.

The identification process also includes different levels of granularity of source code changes. 
In some PSs~\citeS{Sliu2018connecting,Ssharma2015post}, the analysis of code changes only consider the entire file, it is common when it aims to obtain an overview of some property or tendency (eg. code quality).
Some PSs consider code changes in the level of classes~\citeS{Smolnar2017discovering} or just as attributes and methods~\citeS{Skitsu2013detecting,Sdotzler2017more}.
Kitsu et al.~\citeS{Skitsu2013detecting} described code changes based on types of program changes of adding, deleting, moving and changing on the classes, methods, and fields.
The most precise approaches, which were presented in PSs~\citeS{Syuan2013changechecker, Sdotzler2016move, Slin2016empirical, Snguyen2016using, Skaur2016predicting, Sfrick2018generating, Snguyen2013study, Smolderez2017mining, Ssantos2017recommending,Sstevens2017extracting,Smartinez2013automatically,Smeng2013lase,Sosman2014mining,Srolim2017learning,Sjacobellis2014cookbook,Sfalleri2014fine,Snegara2014mining,Shead2017writing,Smaruyama2016slicing,Sdagit2013identifying,Sliu2018closer,Szhong2018towards,Shigo2015toward,Ssong2018systematic,Sphothilimthana2017high}, implement fine-grained differencing techniques, which allows identifying the modifications in syntax terms.

In the PSs~\citeS{Ssilva2017refdiff,Stsantalis2018accurate}, code changes such as refactorings usually are identified by predefined rules, which define properties between code entities to a specific refactoring type.

\subsection{Applying source code changes (RQ2)}
\label{section:rq2}

The performing of source code changes is strict correlated to RQ1 because the modification sequence is defined by the edit script, which is generated in the differencing algorithm.
The PSs~\citeS{Sdotzler2016move,Sfrick2018generating,Sstevens2017extracting,Sfalleri2014fine} perform this process on ASTs, in which the edit script is composed of a set of operations which include inserting, deleting, moving and updating actions, to provide a sequence more understandable and closer to a sequence conducted by the developer.

Usually, source code changes have been achieved by providing examples, as a sequence of structural changes.
Martinez et al.~\citeS{Smartinez2013automatically} evaluated the representation of source code change patterns using before/after AST code hunks in different software repositories.
Meng et al.~\citeS{Smeng2013lase} designed an approach to learning and applying systematics edits by examples, which considers change context to generate the edit script.
In another approach, Rolim et al.~\citeS{Srolim2017learning} have used a domain-specific language to describe these source code changes, which also are deducted and ranked.

The applying of source code changes also has been targeted by specific contexts as refactorings and to specific types of programming languages.
Raychev et al.~\citeS{Sraychev2013refactoring} presented a system of refactoring synthesis based on examples provided by users, to conduct at least a desire sequence of transformations.
Song and Tilevich~\citeS{Ssong2018systematic} also used user examples to perform transformations but in the context of web programming languages.

From a different perspective, Stevens and Roover~\citeS{Sstevens2017extracting} presented an approach to extracting executable transformations based on an evolution query, which describes the sought-after sequence of source code changes.

\subsection{Identifying source code patterns (RQ3)}
\label{section:rq3}

During the development process, the practice of repetitive solutions highlighted the opportunity to reuse previous knowledge in the source code.
Nguyen et al.~\citeS{Snguyen2013study} pointed out the repetitiveness of source code changes and bug-fixes as a twofold opportunity, in the same repository and between repositories.
Higo et al.~\citeS{Shigo2015toward} investigated how often the occurrence and the presence of cross-project code changes.

Some PSs~\citeS{Sdagit2013identifying,Smolderez2017mining,Snegara2014mining} have applied similarity and grouping strategies to identify possible patterns in the development process.
Dagit and Sottile~\citeS{Sdagit2013identifying} identified patterns of code changes extracted from version history using metrics of structural similarity and pattern extraction via antiunification.
Molderez et al.~\citeS{Smolderez2017mining} extracted and grouped source code change to identify unknown systematic edits.
Similarly, Negara et al.~\citeS{Snegara2014mining} presented an approach to the identification of frequent unknown patterns in the code change practices produced by developers.

Other PSs~\citeS{Shigo2014mpanalyzer,Sosman2014mining,Shanam2016discovering,Szhong2018towards,Sliu2018closer} have used code patterns to identify possible bugs and how to better strategies to fix them.
Hugo and Kusumoto~\citeS{Shigo2014mpanalyzer} used the source code patterns extracted in past changes, to identify unintended inconsistencies and incomplete code changes, which may introduce bugs in software.
Osman et al.~\citeS{Sosman2014mining} identified recurrent bug-fixes patterns through explored hundreds of software repositories that have the potential to automatization.
Hanam et al.~\citeS{Shanam2016discovering} have also discovered some relevant bugs patterns but in the specified context of a specified programming language.
Zhong and Meng~\citeS{Szhong2018towards} evaluated the contribution of code structures of past fixes in the automatic program repair process.
Liu et al.~\citeS{Sliu2018closer} identified plenty of opportunities to improve based on real-world patches in different levels of granularity, not only statements.

Besides, some PSs~\citeS{Smaruyama2016slicing,Slin2016empirical} have focused on specific problems.
Maruyama et al.~\citeS{Smaruyama2016slicing} extracted a collection of fine-grained code changes that may be related to a particular program entity based on a recorded change history.
Lin et al.~\citeS{Slin2016empirical} analyzed the characteristics of fine-grained source code change types in dynamic languages, which have considered different projects and versions.

\subsection{Measuring of source code quality (RQ4)}
\label{section:rq4}

Code quality metrics have been used to evaluate software health.
Some PSs~\citeS{yuan2013changechecker,Smolnar2017discovering} have employed metrics to identify defects and relevant changes.
Yuan et al.~\citeS{yuan2013changechecker} used patterns of code changes to extract a collection of quality metrics, to predict the presence of defects during the software development process.
Molnar and Motogna~\citeS{Smolnar2017discovering} analyzed some code quality metrics across different software repositories to observe its behavior, to identify relevant code changes.

Through the continuous analysis of these metrics, the PSs~\citeS{Sislam2018characteristics,Sliu2018connecting,Ssharma2015post,Skaur2016predicting,Snisa2015fault} have elaborated approaches to predict the occurrence of changes and defects and identify them across project versions.
Islam and Zibran~\citeS{Sislam2018characteristics} compared buggy and non-buggy code clones in quality perspective, through analysis of software version history by extracting code quality metrics.
Liu et al.~\citeS{Sliu2018connecting} used software and process metrics to build a historical version sequence of metrics, to predict in file-level defects through applying a recurrent neural network.
Sharma et al.~\citeS{Ssharma2015post} designed four new metrics to understand code change evolution across versions in software repositories.
Kaur et al.~\citeS{Skaur2016predicting} have evaluated whether code smells may be better predictors of change-proneness than static code metrics.
Nisa and Ahsan~\citeS{Snisa2015fault} evaluated the performance of the different machine learning classifiers to elaborate a fault prediction model using code and design metrics.

\subsection{Suggesting of source code changes (RQ5)}
\label{section:rq5}

The process of suggesting source code changes has different points that may be explored.
The PSs~\citeS{Snguyen2016using,Sdotzler2017more} proposed strategies to identify potential locations for applying changes.
Nguyen et al.~\citeS{Snguyen2016using} have used code change patterns to suggest transformations that belong to the same task or context.
Dotzler et al.~\citeS{Sdotzler2017more} observed imprecisions in code recommending mainly related to moving actions, so they proposed a better accuracy approach through the building of code patterns.
To compare strategies, Santos et al.~\citeS{Ssantos2017recommending} evaluated three different approaches (structural, AST-based and Information Retrieval based) to recommending source code locations for specific system transformations.

In another perspective, some PSs~\citeS{Sjacobellis2014cookbook,Skessentini2017context,Snguyen2016api} focused on how to rank code changes.
Jacobellis et al.~\citeS{Sjacobellis2014cookbook} proposed a code completion technique, which recommends the most specific generalization based on the current developer edit stream and a library of the previous edit recipes.
In an industrial study, Marouane et al.~\citeS{Skessentini2017context} proposed an approach of recommending refactorings based on editing context, creating a profile related to recent code changes, fixing bugs and refactorings opportunities, to optimizing its number and reducing antipatterns.
Nguyen et al.~\citeS{Snguyen2016api} developed a statistical model to learning strategies to recommend fine-grained code changes of APIs source code.

At last, there are PSs~\citeS{Shead2017writing,Sphothilimthana2017high} which aim to provide hints to support the applying of changes and bug fixing.
In the course programming context, Head et al.~\citeS{Shead2017writing} presented a mixed system to support the learning process, that fixes the submissions based on previous examples or examples provided by a tutor.
Phothilimthana and Sridhara~\citeS{Sphothilimthana2017high} designed a hint generation system to help students based on different types of misconceptions present in submissions.

\section{Discussion}
\label{section:discussion}

This section discusses the presented results in Section~\ref{section:results}, in order to identify possible practices and open issues.

\subsection{Identifying and applying source code changes (Related to RQ1 and RQ2)}
The process of identifying and applying source code changes has been widely conducted for different purposes.
The majority part of studies has adopted the fine-grained using AST representation of the program transformations, which allows more precise analysis and significative results.

Commonly, each code change is represented by example, a pair of code hunks, one before and another after the specified transformation. Through examples, an edit script is extracted, which represents the sequence of nodes operations performed in the source code.
The current state-of-art differencing technique~\citeS{Sfalleri2014fine} has improved the differencing process.
However, there are left opportunities to become the edit-script shorter and more understandable yet, to get more similar to developer changes.
The PSs~\citeS{Sdotzler2016move,Sfrick2018generating} presented some advances in the mapping task. 
The main problems of these approaches are related to identifying move and update operations, in many cases, these changes are misidentified by sequences of deleting and inserting nodes over the same or different ASTs.
In the edit-script generating task, the optimized algorithm for hierarchical structures proposed by Chawathe~\cite{chawathe1996change} has been adopted.

\subsection{Reusing source code patterns (Related to RQ3)}

Many PSs~\citeS{Shigo2014mpanalyzer,Sosman2014mining,Shanam2016discovering,Szhong2018towards,Sliu2018closer} have pointed out the repetitiveness of some types of transformations.
The main context of these PSs is related to defects, in which each study aims to identify patterns of bugs and its fixes using examples.
Despite the relevant activity, it contemplates corrective code changes.
There is still a huge gap in the investigation between corrective and adaptive source code changes.

The most representative example of adaptive approaches is characterized by software refactoring.
These approaches aim to identify opportunities for applying specified types of refactorings.
However, even the state-of-art approaches~\citeS{Ssilva2017refdiff,Stsantalis2018accurate} have critical limitations, they are twofold: all refactorings are based on the predefined catalog~\cite{refactoringCatalog} and the rate of covered transformations is only 28.8\%~\cite{al2015identifying} to the majority of approaches.

Besides, the identifying of refactorings is currently performed by the definition of rules and heuristics. The applying of examples, in a similar way to the applied with defects, may improve the refactorings activity, to increase the number of contemplated code changes.

\subsection{Quality evaluation in source code changes (Related to RQ4)}

The quality evaluation of source code changes contemplates many applications, since the impact of changes and defects in metrics until complex prediction models.

Some PSs~\citeS{Syuan2013changechecker,Smolnar2017discovering,Sislam2018characteristics,Sliu2018connecting,Ssharma2015post} have evaluated quality metrics about the presence or proneness of defects or relevant code changes.
However, all these studies focus on corrective code changes, and there is no previous study evaluating the impact on quality metrics in adaptive code changes, as refactorings and general improvements.
A possible analysis can identify opportunities for applying specific changes, as a bug prediction model~\citeS{Skaur2016predicting,Snisa2015fault}, but targeting adaptive changes, or at least, it alerts about a degradation process of a particular component or project.

In respect of quality metrics, the majority part of PSs has focused on code metrics, mainly related to complexity and object-oriented.
Liu et al.~\citeS{Sliu2018connecting} used some process metrics with machine learning models, but their results were not relevant.

\subsection{Suggesting code changes (Related to RQ5)}

In summary, the PSs related to suggesting code changes have covered one or more of the following tasks: identify potential locations to applying code changes, rank the most adequate recommending and provide hints related to suggested transformations.

For the first task, the use of AST-based approaches has presented the best results in comparison to structural and information retrieval alternatives~\citeS{Ssantos2017recommending}.
This task also can be supported by additional analysis of features, such as structural context~\citeS{Snguyen2016using,Sdotzler2017more,Sjacobellis2014cookbook} and quality metrics,  to provide more accurate recommendations.

The remaining challenge (related to second and third tasks) is how to integrate the suggestions with appropriate hints about the respective source code changes.
The PSs~\citeS{Shead2017writing,Sphothilimthana2017high} have identified benefits of hint-generating approaches in course programming contexts, in which a tutor or professor evaluated, at least, some submissions of the students and provided the respective fixing and/or hint.
However, both approaches have a supervisor, in a software project context, and there is no specified role to evaluate the quality of development in an industrial project.
Future works can proposed alternatives to provide more complete complimentary feedback in the development process. 
One possible way is the use of examples as the provided pattern, as previously presented approaches, with the addition of personal feedback by developers.

\section{Limitations of this review}
\label{section:limitations}

The systematic review extracted studies from four different search engines, which each one is considered relevant in the study context, however relevant studies may have been missed.
The criteria of study selection have focused on characteristics related to research questions, to obtain the respective answers.
The entire selection process was conducted in pairs, in which an experimenter and reviewer evaluated each one of the studies, to avoid possible mistakes or bias decisions.
We verified the relevant of select studies, through a quality assessment process, which was also performed by an experimenter and reviewer.
To ensure the validity of our interpretations, we consulted additional sources as previous papers as a reference, to improve the understanding of the respective study.

\section{Final Remarks}
\label{section:final-remarks}
We identified 2410 papers from the literature in four search engines, of which 39 were selected as a result of their align to subject and required quality of this review.
The studies were grouped according to each of the research questions, to identify the best outcomes for each point.
We identified that source code changes are used to different subjects, since the differencing process until code change suggesting and hints.
All these tasks can provide support in the development process.
In this way, we highlight the most relevant points and possible trends for each activity.

In summary, we evidence the potential area of search to adaptive source code changes. 
The majority of studies have obtained improvements using source code changes as examples, to corrective purposes, eg. pattern identification of bugs and its fixes respectively.
The studies related to adaptive code changes have focused on refactorings, but in a very limited approach, which use a limited set of options and predefined rules to identification, despite its relevancy.
To achieve advances in this area, researchers may use auxiliary tools as code metrics, past code changes or even developer feedback, to identify the best practices of coding activity and stimulate its replication.


\section*{Acknowledgments}
\label{section:acknowledgement}
We are grateful to CAPES Foundation by the support. 
Our most sincere thanks to the participants from the Laboratory of Software Engineering and Applications (LaPESA) by the support in this review.

\appendix

\section{Studies included in the review}
\label{appendix:included-studies}

\bibliographystyleS{elsarticle-num}
\bibliographyS{studies}

\section{Acceptance and Exclusion Criteria}
\label{appendix:criteria}

This appendix presents the set of criteria elaborated in the systematic review, to support the study's selection.
Tables~\ref{table:acceptance-criteria} and~\ref{table:exclusion-criteria} presents acceptance and exclusion criteria respectively.

\begin{table}[!ht]
\caption{List of acceptance criteria}
\label{table:acceptance-criteria}
\begin{tabularx}{\columnwidth}{p{.8cm}|X}
\toprule
\textbf{ID} & \textbf{Criteria}\\
\midrule
AC 1 & The study is related to the version history in the context of source code changes or code quality\\\hline
AC 2 & The study presents techniques, approaches regarding the detection of source code changes.\\\hline
AC 3 & The study proposes or reports something about automated learning of source code changes based on static analysis.\\\hline
AC 4 & The study is concerned with the learning of source code change patterns.\\\hline
AC 5 & The study is related to measuring code quality in source code changes.\\\hline
AC 6 & The study examines the impact of improvements, defects in source code quality.\\\hline
AC 7 & The study reports the use of metrics for analysis and classification of source code quality.\\\hline
AC 8 & The study presents some strategy of suggestion of source code changes.\\
\bottomrule
\end{tabularx}
\end{table}

\begin{table}[!ht]
\caption{List of exclusion criteria}
\label{table:exclusion-criteria}
\begin{tabularx}{\columnwidth}{p{1cm}|X}
\toprule
\textbf{ID} & \textbf{Criteria}\\
\midrule
EC 1 & The study does not have abstract.\\\hline
EC 2 & The study is only published as a summary or poster.\\\hline
EC 3 & The study is not written in English.\\\hline
EC 4 & The study is an older version of the other study already considered.\\\hline
EC 5 & The study is not a primary study.\\\hline
EC 6 & The study could not be accessed.\\\hline
EC 7 & The study was published before 2012.\\\hline
EC 8 & The study does not present static analysis, but it uses other approaches.\\\hline
EC 9 & The study involves very specific source code changes of a given context.\\\hline
EC 10 & The study is not related to Software Engineering.\\\hline
EC 11 & The study is not related to source code changes.\\\hline
EC 12 & The study is related to source code changes, but it is not intended to identify, learn or analyze their impact.\\\hline
EC 13 & The study involves source code changes, but it is focused on other applications as visualization.\\
\bottomrule
\end{tabularx}
\end{table}

\section{Process of Quality Assessment}
\label{appendix:quality-criteria}
In this section, we present the questionnaire of quality assessment. Each question assigned a score between 0 -- 1 in the study evaluation.

Table~\ref{table:questionnaire} presents the defined questions and their respective points were considered in the process analysis.
\begin{table*}
\caption{Quality Assessment Questionnaire}
\label{table:questionnaire}
\begin{tabularx}{\textwidth}{X}
\toprule
\textbf{1. Is the article based on scientific research?}\\
-- Is the conducted study based on research or it just reports lessons learned from expert opinion in the area?
\\\hline
\textbf{2. Are the goals clearly defined?}\\
-- Does the paper highlight benefits from the achievement of the presented goals?
\\\hline
\textbf{3. The study context was clearly presented?}\\
-- Does the paper identify some problems inside of study context?\\
-- Are previous solutions mentioned and their respective disadvantages?
\\\hline
\textbf{4. Is the approach/strategy developed clearly described?}\\
-- Did the researchers justify the study approach applied?\\
-- Are there other studies that have applied the same study approach?
\\\hline
\textbf{5. Did the researchers analyze the advantages/disadvantages/limitations of the approach/strategy?}\\
-- Does the approach have restrictions or low applicability?
\\\hline
\textbf{6. How much relevant are the research questions?}\\
-- Did the research questions adequately address the study problem?
\\\hline
\textbf{7. Is data study used in real applications or just experimental context?}\\
-- Is the data elaborated just for the conducted study or it is from real application in the industry in the specified context?\\
-- Can the selected data sources influence the study results?
\\\hline
\textbf{8. Were the results obtained and analyzed clearly described?}\\
-- Are there explicit findings in the study?\\
-- How relevant are the identified findings?
\\\hline
\textbf{9. Was the influence of the researchers evaluated on the results?}\\
-- Did researchers identify potential influence points?\\
-- Can these influence points invalidate the results?\\
-- Are the limitation of the study discussed?
\\\hline
\textbf{10. Is there any discussion of the results?}\\
-- Do the researchers have explained the impact of obtained results?\\
-- Are the conclusions justified by the results?
\\\hline
\textbf{11. Were relevant contributions founded in the study?}\\
-- Are the contributions aligned to the tendency of previous studies results?\\
-- How the identified contributions influence in future works?
\\\bottomrule
\end{tabularx}
\end{table*}

\section{Data Extraction Form}
\label{appendix:form-data-extract}

This section presents the form of extraction data applied in the review. The collected data were used for the synthesis of the obtained results. 
Table~\ref{table:data-extraction} presents and describes the collected attributes of each paper.

\begin{table*}
\caption{Data extraction form}
\label{table:data-extraction}
\begin{tabularx}{\textwidth}{l l X}
\toprule
\textbf{ID} & \textbf{Attribute} & \textbf{Description}
\\\midrule
1 & Study identifier & Unique identifier for each study
\\
2 & Extraction date & 
\\
3 & Authors & List of all authors
\\
4 & Year of publication & 
\\
5 & Data sources & Bibliography base of the paper
\\
6 & Study overview & Summarized description of study
\\
7 & Context & It describes the context related to the study problem
\\
8 & Objectives & What were the aims of the study?
\\
9 & Strategy of Approach/Technique & It describes the general steps of the proposed approach or technique
\\
10 & Type of source code change & It identifies the granularity level of source code change
\\
11 & Evaluation process & It presents an overview step of the evaluation process
\\
12 & Results & What were the main results obtained?
\\
13 & Discussion of results & What were the consequences of the results? 
\\
14 & Conclusion and Contributions & What were the contributions and possible future works?
\\\bottomrule
\end{tabularx}
\end{table*}


\section*{References}

\bibliography{references}

\end{document}